# Get 'em Moles! : Learning Spelling and Pronunciation through an Educational Game

**Dhruv Chand**
National Institute of Technology Karnataka
Surathkal, Karnataka, India
dhruvchand@live.com

**Karthik Gopalakrishnan**
Indian Institute of Technology Patna
Patna, Bihar, India
karthik.cs11@iitp.ac.in

**Nisha K K**
National Institute of Technology Karnataka
Surathkal, Karnataka, India
nishakk94@gmail.com

**Mudit Sinha**
Vellore Institute of Technology
VIT University
Chennai, Tamil Nadu, India
mudit.sinha2012@vit.ac.in

**Shreya Sriram**
Delhi Technological University
New Delhi, India
shreyasriram29@gmail.com

## Abstract
Get 'em Moles! is a single-player educational game inspired by the classic arcade game Whac-A-Mole. Primarily designed for touchscreen devices, Get 'em Moles! aims to teach English spelling and pronunciation through engaging game play. This paper describes the game, design decisions in the form of elements that support learning, preliminary play-testing results, and future work.

## ACM Classification Keywords
K.3.1;

## Introduction
Effective verbal communication calls for good pronunciation, and effective written communication requires a good knowledge of spelling. According to the ERIC report [3] 'Why Teach Spelling?', a skilled speller is both a strong reader and a strong writer. Get 'em Moles! (Figure 1) is a single-player game aimed at teaching English spelling and pronunciation through play. We designed and developed a prototype of the game using the Unity 3D game engine.

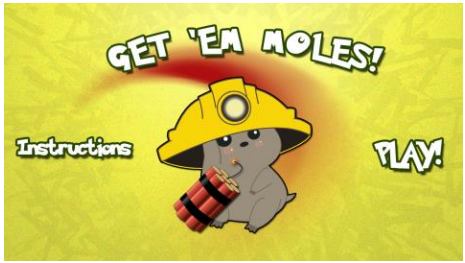

**Figure 1**: Get 'em Moles! start screen

## Objectives

The objective of the game is to teach the right spellings and pronunciations of English words. A basic knowledge of the English language and familiarity with the QWERTY keyboard layout are the prerequisites to play the game. A fixed word list is used in the game. However, the game can be easily extended to use multiple word lists of different levels of difficulty, ranging from commonly used words to advanced words.

## Game Play

The game involves typing the right spelling of each word after it is spoken by the in-game voice and gaining points as reward. Hints are provided by Murphy the Mole in the form of bombs planted by Murphy under the letters of the on-screen keyboard.

The main game screen (Figure 2) comprises of a stylized QWERTY keyboard, score meter, pause/play button, a 'Play Again' speaker button to listen to the pronunciation of a word again, and a word text area that displays letters of the word that have already been typed in. The in-game voice speaks a word from the word list, after which the game waits for the player to hit the letters corresponding to the word's spelling on the on-screen keyboard, one-by-one.

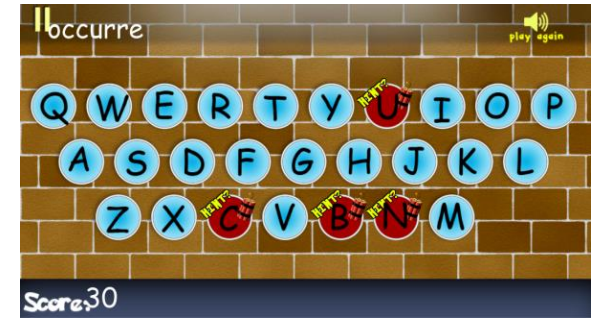

**Figure 2**: Main game screen showing a stylized QWERTY keyboard and other components

If the player hits the correct letter, the letter turns green, a pleasant chime sound is played and the letter appears in the word text area above. The letter is also spoken by the in-game voice after it is hit for additional positive reinforcement.

If the player does not hit any letter within a fixed period of time, some letters on the keyboard glow red and small bombs appear near them. One of these letters is correct and the rest are decoys. The decoys are selected randomly. This provides the player a choice of possible answers for the next letter in the word as a hint. If the player hits the correct letter now, the player gets lesser points when compared to hitting the correct letter without the hint. If the player does not hit any letter, the bombs explode after a fixed period of time, revealing Murphy the Mole under the correct letter (Figure 3). The player is then left with no choice but to hit the correct letter and move on to the next letter of the word. This does not earn the player any points.

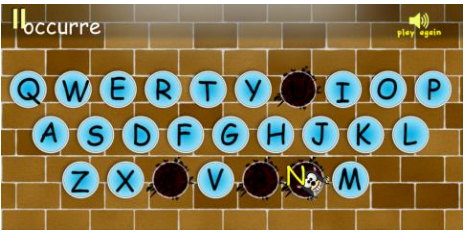

**Figure 3**: Giveaway hint revealing the next letter for the word "occurrence"

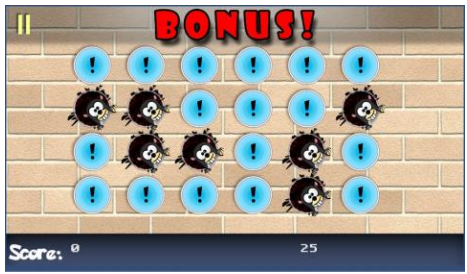

**Figure 4**: Whac-A-Mole (Bonus Round)

At any stage, a wrong hit causes a buzzer to sound and the hit letter turns red to indicate that it is the wrong letter. The correct letter is also revealed through Murphy the Mole's appearance under the correct letter. A high score streak invokes a bonus round (Figure 4), where players get to play Whac-A-Mole and increase their overall score. The bonus round is time-limited and fast-paced. A video of the game being played on a tablet is available at the following link: vimeo.com/132009027

### Game Elements Supporting Learning

By using a combination of visual and audio signals when both correct and incorrect letters are hit, the game enables effective learning by providing known benefits of multisensory learning [5]. Immediate feedback on performance [1] is also incorporated through the same signals and hints, which is known to make learning more effective. The game features that support learning are as follows:

- The pronunciation of a word can be played multiple times using the 'Play Again' button.
- Every correct hit is supported by the letter turning green and the pleasant sound of a chime. This is supplemented by the letter being spoken by the in-game voice and displayed in the word text area on the top of the screen for further cognitive reinforcement.
- Every incorrect hit is supported by the letter turning red and the sound of a buzzer. The correct letter is provided as a giveaway hint to immediately learn after making the mistake.
- A multiple-choice hint and eventually a giveaway hint revealing the correct letter is provided when the player doesn't hit any letter within a given delay, presumably since the player doesn't know the correct letter. This is to enable the player to learn and progress further.

### Play Testing and Results

We randomly selected ten $5^{th}$ and $6^{th}$ grade students from a school in Bangalore, India, and conducted a preliminary play-testing session with them. The objective of the session was to see whether the students learn the right spellings of words by playing the game. We also obtained feedback on the game from the students through interviews.

We first conducted a pre-test (Figure 5), asking the students to write, on separate sheets of paper, the spellings of 5 words dictated to them. After collecting their papers, we asked each student to play the game, which featured a superset of the set of words from the pre-test, for about 15 minutes each on a touchscreen device. After playing the game, the students took a post-test, in which the same 5 words from the pre-test were dictated to them and they had to write down the spellings of the words on separate sheets of paper. We then evaluated both sets of test papers and looked for students who misspelled a word before playing the game and spelled the word correctly after playing the game. We found that 4 students belonged to this category. Additionally, we observed that no student misspelled a word both before and after playing the game.

In informal interviews conducted after we evaluated all the test papers, all the students reported enjoying the game and said they would definitely like to play the game again. The 4 students we identified previously were individually informed that their answers for the same word were different in the two tests they took.

We asked each of them which of the two answers they believed to be correct. All of them pointed to their answers from the post-test and said they learned the correct spellings of those words through the game. Hence, we conclude from the observed instances of improvement and from the interview feedback that the game enables learning in practice and in an enjoyable manner.

## Future Work

Having seen that students learn the right spellings of words by playing the game, we will be conducting play-testing sessions to see whether students also learn the right pronunciations of words by playing the game.

We plan on undertaking a formal evaluation with a full version of the game to better assess its learning potential. We would like to find out if the game is more effective than traditional methods of learning the right spellings of words. The development of a full version is currently in progress. The full version includes many word lists of different levels of difficulty. The technique of spaced repetition [2] is being incorporated and we also plan to make the game adaptive, which has been shown to lead to better learning outcomes [4]. This would allow players to automatically graduate to words of greater difficulty depending on their performance.

We foresee a wide range of audiences for the game, from kids preparing for the Scripps National Spelling Bee to undergraduates preparing for the Graduate Record Examinations (GRE), an essential component of graduate school applications. We hope to learn about the efficacy of the game across a range of audiences by releasing the full version online for a variety of platforms upon completion of development.

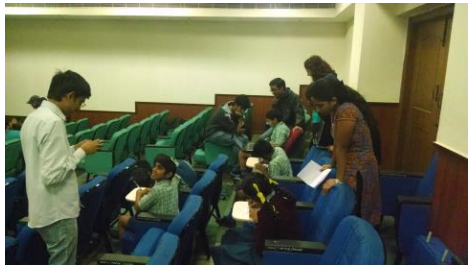

**Figure 5**: Play-testing session with some of the students

## Acknowledgements

We acknowledge Whac-A-Mole as the source of inspiration for this game. We thank Amy Ogan, Erin Walker and Erik Harpstead for their guidance in developing and evaluating Get 'em Moles!. We thank MS Ramaiah Institute of Technology, Bangalore, India, for enabling us to play-test our game with children. Finally, we thank Carolyn P. Rosé and Carnegie Mellon's Internship Program in Technology Supported Education, which brought us together for 15 days to conceptualize, develop and undertake a preliminary evaluation of the prototype for this game.